\documentclass[sigconf,screen]{acmart}
\usepackage{graphicx}
\usepackage{setspace}               % for LINE SPACING
\usepackage{multicol}               % for MULTICOLUMNS
\usepackage{xcolor}
\usepackage{caption}
\usepackage{subcaption}
\usepackage{csquotes}

\usepackage[utf8]{inputenc}
\usepackage{url}
\usepackage{hyperref}
\usepackage{amsfonts}
\usepackage{float} %For [H] in figures
\usepackage{listings}
\usepackage{color}
\usepackage{fancyhdr} %For headers http://tex.stackexchange.com/questions/118647/line-under-chapter-name-at-top-of-the-page
\usepackage{bibentry} 
\usepackage{enumitem}
\usepackage{multirow}
\usepackage{geometry}
\usepackage{dirtytalk}
\usepackage{tabularx}
\usepackage{balance}

\usepackage[english]{babel}
\usepackage{soul}

\definecolor{edit}{HTML}{000000}

\fancypagestyle{firstpage}
{
    \fancyhead[L]{Workshop paper presented at XAIxArts'25 – the \href{https://xaixarts.github.io/2025}{third international workshop on e\textbf{X}plainable \textbf{AI} for the \textbf{Arts}}, held in conjunction with the \href{https://cc.acm.org/2025/}{ACM Creativity and Cognition conference 2025}, June 23rd, 2025.}    
    \fancyhead[R]{}
}

\AtBeginDocument{%
  \providecommand\BibTeX{{%
    \normalfont B\kern-0.5em{\scshape i\kern-0.25em b}\kern-0.8em\TeX}}}

\setcopyright{none}           % ← turns the permissions block off
\acmConference[\href{https://xaixarts.github.io/2025}{XAIxArts 2025}]{Explainable AI for the Arts Workshop 2025}{June 23, 2025}{Online}
\acmDOI{}      % no DOI line
\acmISBN{}     % no ISBN line
\acmPrice{}    % no price line
\begin{document}

%\title[Shortened Paper Title]{Outcomes and Creativity Support Tools: Aiding the Output or Aiding the User Themselves?}
%\title{What are valuable measures when studying GAI for the arts?}  
%\title{Evaluating XAI using measures of user-centric benefits}  
%\title{Should my AI explanations be evaluated for emotional well-being, self-reflection, (or other user-centric benefits)?}  
\title{Reflecting Human Values in XAI: Emotional and Reflective Benefits in Creativity Support Tools}

\author{Samuel Rhys Cox}
\email{srcox@cs.aau.dk}
\orcid{0000-0002-4558-6610}
\affiliation{%
  \institution{Aalborg University}
  \city{Aalborg}
  \country{Denmark}
}

\author{Helena Bøjer Djernæs}
\email{hbd@cs.aau.dk}
\orcid{0009-0003-9457-811X}
\affiliation{%
  \institution{Aalborg University}
  \city{Aalborg}
  \country{Denmark}
}

\author{Niels van Berkel}
\email{nielsvanberkel@cs.aau.dk}
\orcid{0000-0001-5106-7692}
\affiliation{%
  \institution{Aalborg University}
  \city{Aalborg}
  \country{Denmark}
}

\begin{abstract}
%In this workshop paper, we discuss alternative outcome variables that could be measured and explored when evaluating explainable AI (XAI) systems. As a background to this, we draw from our recent review of creativity support tool (CST) evaluations which found a paucity of studies evaluating CSTs using user-centric measures that benefit the user themselves. Specifically, measures of developing intrinsic abilties, emotional well-being, self-reflection, and self-perception.
%On from this, we discuss these user-centric measures within the context of XAI, and wish to provoke and discuss the potential for of such measures with XAI and the arts.
In this workshop paper, we discuss the potential for measures of \textit{user-centric benefits} (such as emotional well-being) that could be explored when evaluating explainable AI (XAI) systems within the arts. 
As a background to this, we draw from our recent review of creativity support tool (CST) evaluations, that found a paucity of studies evaluating CSTs for user-centric measures that benefit the user themselves.
Specifically, we discuss measures of: (1) developing intrinsic abilities, (2) emotional well-being, (3) self-reflection, and (4) self-perception.
By discussing these user-centric measures within the context of XAI and the arts, we wish to provoke discussion regarding the potential of such measures.

\end{abstract}

\maketitle

\thispagestyle{firstpage}

\section{A Brief Background and Motivation}

Explainable AI (or XAI) is a field focused on making the outputs and inner workings of AI models more interpretable and understandable to human users~\cite{10.1145/3290605.3300831}.
This could be via visual~\cite{10.1145/3706598.3714004,10.1145/3313831.3376615} or text-based~\cite{10.1145/3706598.3713730,10.1145/3686922} explanations of AI models.
%For this explanations can be provided to users, such as visual or text-based explanations of AI models~\cite{10.1145/3290605.3300831}.
%This could be by the provision of (perhaps more data heavy and esoteric) explanations to AI engineers to explain 
%For example, explanations could be used to help people understand and choose between multiple AI recommendations~\cite{10.1145/3706598.3713730,10.1145/3313831.3376615}.
For example, an XAI system could explain a decision in relation to a multitude of complex impacting variables (such as visualising physiological measures affecting a health diagnosis~\cite{10.1145/3290605.3300831}, or factors affecting the price of a home~\cite{10.1145/3313831.3376615}).
This can be applied to domains such as stress coping strategies, where, for example, text-based counterfactual explanations~\cite{10.1145/3706598.3713730} can be provided to help users select a preferred stress coping strategy.

However, the provision of AI explanations has historically been more productivity-centred~\cite{bryan2025XAIxArts,rong2023towards}, which may conflict with the purpose of the arts where one may want to focus on improving aspects to the \textit{user themselves} such as increased feelings of empowerment, enjoyment, and ownership of creative artefacts.
The feeling of AI impeding these creative ideals has been found in multiple domains (such as creative writing~\cite{10.1145/3532106.3533506,cox2025beyond}) where people receiving AI support wish to retain feelings of control, autonomy, and ownership during the creative process.
%This comes to a head in creative domains (e.g., creative writing~\cite{10.1145/3532106.3533506,cox2025beyond}) where writers receiving AI support want to retain feelings of control, autonomy, and ownership during the creative process.
Additionally, the arts have been shown to improve both people's physical health and well-being (see a scoping review of over 3000 papers by Fancourt and Finn for a very clear and broad discussion of this~\cite{fancourt2020evidence}).

Given this background, in this workshop paper we discuss how explanations from XAI systems could be evaluated to focus on outcomes that benefit the \textit{users themselves}, such as emotional well-being or self-reflection. 
This is contrast to perhaps ``typical'' evaluations of XAI systems, that may deploy more productivity-focused measures.
%This goes beyond the productivity focus of prior XAI evaluations, that may focus more on (perhaps as expected for explanations) the quality of explanations in terms of understandability.
For example, Rong et al. reviewed measures used to (empirically) evaluate XAI systems, and found that prior work measured explanations in terms of trust, understanding, usability, and human-AI collaboration performance~\cite{rong2023towards}.
%While these measures certainly make sense (as one would expect explanations to be understandable and trustworthy) we wish to draw discussion to the potential for alternative measures of user-centric benefits.
While these measures certainly make sense 
(as one may expect understandable and trustworthy explanations)
%(as one would expect explanations to be understandable and trustworthy) 
we wish to discuss the potential for measures of benefits to the \textit{user themselves} such as emotional well-being.
%Previous review of human-centred XAI found that studies measured explanations in terms of trust, understanding, usability, and human-AI collaboration performance~\cite{rong2023towards}.
%While these outcomes do sound like the most instantly apparent and sensible points of focus for explanations, it does overlook the same...

%In a manifesto for XAI and the arts, Bryan-Kinns et al. call for attention to four core themes of:  1) Empowerment, Inclusion, and Fairness; 2) Valuing Artistic Practice; 3) Hacking and Glitches; and 4) Openness~\cite{bryan2025XAIxArts}.

This motivation comes from our review of measures used in the empirical evaluation of creativity support tools (CST)~\cite{cox2025beyond}.
%This motivation is drawn from our own publication, where we reviewed measures used in the empirical evaluation of creativity support tools (CST)~\cite{cox2025beyond}.
%Here we reviewed ACM DL publications (2015–2024), and found that (while most studies evaluated user experience or creative artefact quality) only 15\% of studies evaluated CSTs using user-centric measures.
Here, our review of ACM DL publications (2015–2024) found that, although most studies evaluated user experience or creative artefact quality, only 15\% assessed CSTs using user-centric benefits (i.e., developing intrinsic abilities, emotional well-being, self-reflection, and self-perception).
%such as user well-being or self-reflection.
%This paucity of measures of user-centric benefits, meant that outcomes (i.e., developing intrinsic abilities, emotional well-being, self-reflection, and self-perception) were evaluated in only a minority of studies.
%On from this, in a review of ten years of ACM DL studies that evaluate creativity support tools (CSTs), we found a paucity of studies that evaluated tools with ``\textit{user-centric}'' measures (such as emotional well-being)
%We highlight these user-centric outcomes, contextualise them within XAI and the arts, and discuss future directions for CSTs—beyond productivity as in prior work~\cite{bryan2025XAIxArts}.
%We highlight these user-centric outcomes, contextualise them within XAI and the arts, and discuss future directions for XAI in CSTs (that think beyond productivity similar to prior work~\cite{bryan2025XAIxArts,cox2025beyond}).
We highlight these user-centric outcomes, contextualise them within XAI and the arts, and discuss future directions for XAI in CSTs beyond productivity~\cite{bryan2025XAIxArts,cox2025beyond}).
%and (similar to prior work~\cite{bryan2025XAIxArts}) offer a motivation to look beyond productivity as a focus of CSTs
%We highlight these user-centric outcomes that have been somewhat overlooked in prior evaluations of CSTs, and contextualise these against XAI and the arts. Here, we offer discussion for potential future outlooks for the development of XAI, and (similar to prior work~\cite{bryan2025XAIxArts}) offer a motivation to look beyond productivity as a focus of CSTs.

%While the studies surveyed in our review do not explicitly use XAI, 

%Both explainability for expert audiences (such as software developers or data analysts) or novice users

%Both explainability for those \textit{creating} AI (e.g., software engineers) where explanations may focus on making AI fairer and less biased, or explainability as part of the user interaction itself where explanations may help to explain AI decisions.
% Bit shit:
%This can be transferred to art where people may in a sense act as both those creating and making decisions.

%\textit{Note:} Caveat that focus may be more on CSTs generally and to a ``general audience'' rather than framed from the viewpoint of professional or full-time artists (as we are not artists).

%Previous review of human-centered XAI found that studies measured explanations in terms of trust, understanding, usability, and human-AI collaboration performance~\cite{rong2023towards}.
%While these outcomes do sound like the most instantly apparent and sensible points of focus for explanations, it does overlook the same...

\section{What Measures of User-Centric Benefits Could We Use?}

We frame our discussion of user-centric benefits around measures that emerged in our review of CST evaluations~\cite{cox2025beyond} (see \cite[§4.3]{cox2025beyond} for further prior‐work).
%(see~\cite[Section~4.3]{cox2025beyond} for detailed descriptions of prior work that are not contained in this workshop paper).
%\footnote{While (in interests of space) we cannot cite broad examples of prior CST evaluations that deployed certain measures, please see~\cite[Section~4.3]{cox2025beyond} for more detailed and descriptive summaries of prior work.}. 
Further, this workshop paper is differentiated against our review, as our review did not contain or discuss applications to XAI.
%to differentiate this workshop paper against our review, we 
Additionally, a limitation of this paper is that we cannot claim to have reviewed all existing measures using XAI in the arts.
However, while this will in no way be exhaustive of \textit{all} potential measures of user-centric benefits, we hope to spur discussion and potentially motivate focus of future XAI studies.

%Below, we highlight measures of user-centric benefits, namely: (1) Developing Intrinsic Abilities, (2) Emotional Well-being, (3) Self-Reflection, and (4) Self-Perception. These outcomes have received limited empirical evaluation within CST evaluations, and we wish to frame these within the context of XAI and the arts to spur future work.
We highlight four user-centric benefits: (1) Developing Intrinsic Abilities, (2) Emotional Well-being, (3) Self-Reflection, and (4) Self-Perception. These outcomes have seen limited empirical evaluation in CST, and we frame them in XAI and the arts to spur future work.

\vspace{1.25mm}
%\subsubsection*{Developing Intrinsic Abilities}
\noindent
\underline{\textit{\textbf{Developing Intrinsic Abilities:}}}
When evaluating XAI for CSTs, the development of intrinsic abilities, such as learning concepts and terminology, or improving creative skills could be measured.
%XAI could be used to help people develop their intrinsic abilities, such as learning concepts and terminology, and (more specific to the arts) improving creative skills.
Connecting learning and XAI may seem most apparent where one may expect a detailed explanation (that is transparent about an AI's workings or recommendations) to help users form better mental models and learn more effectively.
%This connection between learning and explainability would perhaps be more apparent to readers, where an AI tool that is transparent about its workings or recommendations could help users form better mental models and learn more effectively.
%Further, CST studies measuring developments of intrinsic abilities typically use pre- and post-tests to measure changes in abilities~\cite{cox2025beyond}, and would perhaps necessitate use of validated measures and experiment procedures.
%Contrasting this however, may be a perspective that explanations should help user \textit{discover} a solution, and that XAI should scaffold (e.g., provide hints, ask probing questions, or offer appropriate resources).
Contrasting this however, may be a perspective that XAI should should scaffold (e.g., provide hints, ask probing questions, or offer appropriate resources) to help users \textit{discover} solutions.
%This mindset may come forth given findings that generative-AI may reduce critical thinking~\cite{lee2025impact}.
Findings that generative-AI may reduce critical thinking could enforce this~\cite{lee2025impact}.

Within prior work, providing different visual explanations has been found to aid medical students when training to administer ultrasound~\cite{10.1145/3706598.3714004}.
In our review of CST evaluations~\cite{cox2025beyond}, studies measuring developments of intrinsic abilities typically use pre- and post-tests to measure change~\cite{cox2025beyond}, and such measures and experiment procedures would likely be necessary in XAI evaluations.
%would perhaps necessitate use of validated measures and experiment procedures.
For example, Alves et al. used validated measures and procedures to evaluate improvements in creative skills~\cite{alves2020creativity}.
%Within CST evaluations, studies have used validated measures and procedures to evaluate improvements in creative skills (e.g.,~\cite{alves2020creativity}).
Drawing these together, measuring the impact of XAI on development of intrinsic abilities would provide seemingly fertile ground for investigation, with possibility to measure improvements in learning and creative skills specific to the arts.

\vspace{1.25mm}
%Notably, the connection between learning and explainability is strong. 
%When an AI tool is transparent about its workings, users can form better mental models and learn more effectively
\noindent
\underline{\textit{\textbf{Emotional Well-being:}}}
%\subsubsection*{Emotional Well-being}
Creative activities are often pursued for their emotional and psychological rewards (think the therapeutic effects and joy of creating~\cite{fancourt2020evidence}), and AI tools should ideally amplify, not dampen, these benefits.
Therefore a potential avenue of XAI and the arts, would be to develop explanations that are designed to improve people's emotional well-being. 
Here, XAI could: be emotionally-positive in \textit{style} and \textit{content}~\cite{li2023systematic}, focus on autonomy and empowerment (moderating well-being), and avoid negative comparisons and framing (e.g., explanations could deter perfectionism bias: ``\textit{Please remember: reference images are guides, not rules!}'').
%in XAI framing.
%This could involve changes to explanation \textit{style} and \textit{content} to improve well-being~\cite{li2023systematic}, focusing on autonomy and empowerment (potentially improving emotional well-being), .
%Avoid negative comparisons Framing of XAI (suggestions and not unobtainable gold standard - ``Perfectionism Bias'')

Within CSTs, Gonçalves et al.~\cite{gonccalves2015you} found that creative writing support tools helped improve the well-being of marginalised youth in a two-week longitudinal study.
Within XAI, a recent study found that counterfactual explanations can be given to help people choose coping strategies that lead to reductions in stress (using a validated measure of stress)~\cite{10.1145/3706598.3713730}.

%Creative activities are often pursued for their emotional and psychological rewards – the joy of creating, self-expression, and personal meaning – and (outside of prior literature's focus on productivity~\cite{cox2025beyond}) AI support tools should ideally amplify, not dampen, these benefits.

\vspace{1.25mm}
%\subsubsection*{Self-Reflection}
\noindent
\underline{\textit{\textbf{Self-Reflection:}}}
XAI could be designed to promote people's self-reflection.
Within the arts and HCI, validated measures have been developed to measure reflection as a result of creative activities. 
For example, prior CST evaluations have used the creativity-specific ``\textit{Reflection in Creative Experience}'' (RiCE) questionnaire~\cite{ford2023towards}, and the more general ``\textit{Technology-Supported Reflection Inventory}'' (TSRI)~\cite{bentvelzen2021development}.
%\footnote{See descriptions of the Reflection in Creative Experience (RiCE) questionnaire and Technology-Supported Reflection Inventory (TSRI) in~\cite[Section~4.3.3]{cox2025beyond}.}. 
Further, prior work has investigated whether interactions with CSTs led to users learning new things about themselves, or ideas to overcome challenges as a result of interactions.
From this, a natural progression could be the design of explanations to spur self-reflection beyond explaining the workings of AI models.

For example, Yan et al. developed NaCanva~\cite{yan2023nacanva} to help children develop mood boards to aid in personal reflection. Here, they measured reflection from a variety of novel perspectives such as children's acting and caring on nature, awareness and emotional ties to nature, and feelings and identity within nature. Further, recent work investigated the impact of context and modality (text, image, audio, and video) on self-reflection~\cite{yeo2025enhancing} demonstrating a potential for XAI modality to be explored for self-reflection.

\vspace{1.25mm}
%\subsubsection*{Self-Perception}
\noindent
\underline{\textit{\textbf{Self-Perception:}}}
Finally, XAI systems could be designed to enhance people's self-perceptions, such as self-efficacy and sense of achievement.
On from this, AI systems that are more interactive and explainable could \textit{invite the user} into the loop (so the process still feels like their creative journey), rather than a one-click solution that makes the user feel superfluous.
Additionally, if users understand what the AI is doing and can guide it, they may retain a stronger sense of ownership and achievement.
%AI systems that are more interactive and explainable, inviting the user into the loop (so the process still feels like their creative journey), rather than a one-click solution that makes the user feel superfluous.

Motivation behind these self-perceptions has received more attention within creativity~\cite{10.1145/3613904.3642492,wu_human-generative_2025}. For one example, AI writing support has been found to improve performance but lower intrinsic motivation~\cite{wu_human-generative_2025}.
Additionally, the \textit{style} of explanations can impact self-perception. For example, when completing learning tasks alongside conversational AI feedback, the tone of AI feedback can improve people's self-efficacy~\cite{kollerup2025enhancing}.

%Explanations for children~\cite{wu_human-generative_2025}.

%Both the content of explanations, and the \textit{style} of explanations.

% First ChatGPT conversation for user-centric XAI: https://chatgpt.com/share/681399c9-4eec-800c-a6f9-b9933a64ee67

% Second conversation: https://chatgpt.com/share/681399f5-bc30-800c-8a14-4268a08b24e4

%\section{Discussion and Conclusion}
\section{The Interplay of User-Centric Benefits}

To some extent, the user-centric benefits discussed above are inherently connected.
%For instance, XAI to empower users would directly impact self-perception, while also potentially benefiting emotional well-being~\cite{fancourt2020evidence}.
%For instance, an explanation that centres the user in suggestions and helps to \textit{empower} them (e.g., ``\textit{Your last five painting sessions have used warm hues --- this suggested palette} [...]'') \textit{could} directly benefit self-perception as well as emotional well-being.
For instance, when providing an AI suggestion, XAI could centre the user to help \textit{empower} them, such as the explanation: ``\textit{Your last five painting sessions have used warm hues --- this suggested palette} [...]''. 
%This explanation could then directly benefit people's self-perception, while also mediating people's emotional well-being.
Such an explanation can directly improve the user’s \textbf{self‐perception} and, in turn, enhance their \textbf{emotional well‐being}.
%Similarly, (without gamifying creativity sessions) XAI could focus on past user achievements to both aid emotional well-being or self-reflection (by causing people to reflect on previous creativity). For example, after sharing a suggestion a (in this case, somewhat direct) explanation could be provided: ``\textit{Your last three storyboards used high-contrast lighting to powerful effect. Take a moment to reflect on that success to boost your emotional well-being}''
Similarly, without gamifying creativity sessions, XAI could focus on past user achievements to foster \textbf{self-reflection} and \textbf{emotional well-being}. For example, after offering a suggestion, an explanation could include:
``\textit{Your last three storyboards used high-contrast lighting to powerful effect. Take a moment to reflect on that success to boost your emotional well-being}''.
%As a final example, XAI could help people both learn and thereby gain self-confidence by structuring explanations to be more conversational, or to omit information (cf. studies finding people are more comfortable expressing emotions to chatbots when less feedback is given~\cite{park2021wrote,cox2025ephemerality}).
As a final example, XAI could help people \textbf{learn} and gain \textbf{self-confidence} by structuring explanations to be more conversational or omitting certain details (cf. studies showing that people are more comfortable expressing emotions to chatbots when they receive less feedback~\cite{park2021wrote,cox2025ephemerality}).

%As a final example, XAI could encourage people to pause to think, could behave more conversationally, or incorporate different provisions of information.

%Within CUIs, people may be more comfortable expressing themselves when given less feedback~\cite{park2021wrote,cox2025ephemerality}.

%Take an example of wanting to improve people's feelings of self-expression, which has been shown to improve emotional well-being

%\begin{itemize}
    %\item Self-reflection: Explanations encouraging people to pause. Can have more conversational explanations, or different levels of provision of information.
    %\item People can be more comfortable expressing themselves emotionally when giving less feedback~\cite{park2021wrote,cox2025ephemerality}
    %\item (cf. studies finding people are more comfortable expressing emotions to chatbots when less feedback is given~\cite{park2021wrote,cox2025ephemerality})
    %\item (Without gamifying art too much) XAI could focus on user achievements
    %\item Allow for user control to maintain feelings of ownership.
    %\item Use XAI to centre the user in suggestions, helping to empower them (e.g., ``\textit{Your last five painting sessions have used warm hues --- this palette introduces a cool‐tone option to balance your style.}''
%\end{itemize}

In conclusion, we have provided an overview of several user-centric benefits that could be the focus of future evaluations of XAI systems within the arts.
These benefits, could both shape the evaluation of XAI tools, as well as the design of explanations used themselves.
From this, we hope to spur discussion and motivate fellow researchers to focus on evaluations of XAI tools that focus on user-centric benefits, such as emotional well-being and self-reflection.

%~\cite{10.1145/3613905.3650929}

\begin{acks}
We would like to thank the workshop co-chairs \href{https://orcid.org/0000-0002-6895-2441}{Corey Ford} and \href{https://orcid.org/0000-0001-6212-6627}{Elizabeth Wilson} for organising the workshop, and committee members for detailed reviews that helped improve this paper.
%We would like to thank our reviewers for their positive reception of the work, and their helpful and construction feedback to improve the paper.
This work was supported by the Carlsberg Foundation, grant CF21-0159.
\end{acks}

\balance
\bibliographystyle{ACM-Reference-Format}
\bibliography{sample-base}

% \appendix
% \input{07-Appendix}

\end{document}